# Ultrafast Electron Microscopy of Nanoscale Charge Dynamics in Semiconductors

Michael Yannai[1], Raphael Dahan[1], Alexey Gorlach[1], Yuval Adiv[1], Kangpeng Wang[1], Ivan Madan[2], Simone Gargiulo[2], Francesco Barantani[2,3], Eduardo J. C. Dias[4], Giovanni Maria Vanacore[5], Nicholas Rivera[6], Fabrizio Carbone[2], F. Javier García de Abajo[4,7], Ido Kaminer[1]*

[1]Technion - Israel Institute of Technology; Haifa 3200003, Israel.
[2]Institute of Physics, École Polytechnique Fédérale de Lausanne; Station 6, Lausanne, 1015, Switzerland.
[3]Department of Quantum Matter Physics, University of Geneva; 24 Quai Ernest-Ansermet, 1211, Geneva 4, Switzerland.
[4]ICFO-Institut de Ciencies Fotoniques, The Barcelona Institute of Science and Technology; 08860 Castelldefels (Barcelona), Spain.
[5]Department of Materials Science, University of Milano-Bicocca; Via Cozzi 55, 20121, Milano, Italy.
[6]Massachusetts Institute of Technology; Cambridge MA 02139, USA.
[7]ICREA-Institució Catalana de Recerca i Estudis Avançats; Passeig Lluís Companys 23, 08010 Barcelona, Spain.
*Corresponding author. Email: kaminer@technion.ac.il

**Abstract:** The ultrafast dynamics of charge carriers in solids plays a pivotal role in emerging optoelectronics, photonics, energy harvesting, and quantum technology applications. However, the investigation and direct visualization of such non-equilibrium transport phenomena remains as a long-standing challenge, owing to the nanometer-femtosecond spatio-temporal scales at which the charge carriers evolve. Here, we propose and demonstrate a novel interaction mechanism enabling nanoscale imaging of the femtosecond dynamics of charge carriers in solids. This imaging modality, which we name charge dynamics electron microscopy (CDEM), exploits the strong interaction between terahertz (THz) electromagnetic near fields produced by the moving charges and synchronized free-electron pulses in an ultrafast scanning transmission electron microscope. The measured free-electron energy at different spatio-temporal coordinates allows us to directly retrieve the THz near-field amplitude and phase, from which we reconstruct movies of the generated charges by comparison with microscopic theory. The introduced CDEM technique thus allows us to investigate previously inaccessible spatio-temporal regimes of charge dynamics in solids, for example revealing new insight into the photo-Dember effect, showing oscillations of photo-generated electron-hole distributions inside a semiconductor. Our work lays the foundation for exploring a wide range of previously inaccessible charge-transport phenomena in condensed matter using ultrafast electron microscopy.

**One-Sentence Summary:** Determination of nanometer-femtosecond electron-hole dynamics in solids by direct ultrafast inelastic-electron-microscopy of their associated THz near fields.

At the heart of many solid-state physics phenomena of current interest is the intriguing ultrafast dynamics of charge carriers and their various hybrid states. Novel approaches for imaging ultrafast charge dynamics (*1*) aim to provide new insights in diverse fields such as superconductivity, magnetoresistance, and plasma formation. Many of these phenomena occur on a picosecond timescale, owing to the underlying physics, which involves THz-frequency resonances, as well as meV binding and quasiparticle energies. Imaging the ultrafast charge dynamics involved in these phenomena necessitates powerful microscopy capabilities that combine a sufficient resolution in space (nanometer-scale) and time (sub-picosecond-scale) along with other observables such as energy and momentum.

Here, we propose and demonstrate a new interaction mechanism that enables imaging of ultrafast charge dynamics by inelastic scattering of free-electron-pulse probes in an electron microscope. The inelastic interaction between electron pulses and charge carriers is mediated by time-dependent THz-frequency near fields emitted following picosecond timescale motion of the charge carriers in the bulk of a photoexcited material. This interaction enables the reconstruction of the charge dynamics with nanometer spatial resolution and sub-picosecond temporal precision without perturbing it by the external free-electron probe. We present the first experimental demonstration of this concept and develop a broadly applicable theory in excellent agreement with our measurements.

**Imaging ultrafast charge dynamics**

Currently available microscopy techniques for imaging nanoscale charge dynamics are mostly surface-oriented (i.e., sensitive only to the outermost atomic layers of the target material (*2–13*)), relying on either optical excitation and detection (*2–5*) or analysis of electron emission from the surface (*6–13*). Several specific implementations rely on measurements of THz near-fields scattered by or generated from the surface, enabling the reconstruction of picosecond and sub-micron charge dynamics (*14–16*). Complementary approaches include ultrafast electron diffraction (UED) (*17–20*), which provides momentum-space information on charge dynamics in the bulk; ultrafast point-projection electron microscopy with a spatial resolution of tens of nanometers using low-energy electrons (*21–23*); and mapping of static charge and field distributions from elastic electron scattering (*24*, *25*). Furthermore, charge dynamics associated with the linear optical response of micron-scale metallic resonators has been probed under external THz irradiation using ultrafast electron deflectometry (*26*). Such advances emphasize the potential of free-electron pulses to explore charge-carrier dynamics.

Ultrafast electron imaging experiments are made possible by the invention and swift development of the ultrafast transmission electron microscope (UTEM) (*27*, *28*) – a pump-probe setup that uses femtosecond laser pulses to excite a sample and synchronized femtosecond electron pulses to measure the resulting transient state. Previous studies have utilized UTEMs to image the dynamics of electromagnetic fields in various nanoscale and low-dimensional structures, observing the evolution of plasmons in metallic nanostructures (*29*) and buried interfaces (*30*), confined modes in optical cavities (*31*, *32*), phonons (*33*) and polariton wavepackets in 2D materials (*34*), as well as phase transitions (*35*, *36*) and Skyrmion dynamics (*37*).

These previous UTEM experiments relied on the interaction of the electron beam with optical fields at the light excitation frequency, to which the investigated specimen responded linearly. In contrast, in this work we present a direct demonstration of a strongly nonlinear optical interaction within the UTEM. Specifically, our experiment measures the THz field created by moving charge carriers, which are generated via optical-frequency excitation of a bulk InAs crystal – the so-called photo-Dember effect (*38*). We thus demonstrate free-electron-based near-field imaging of an in-situ generated THz field, revealing new information about the THz generation process and elucidating its spatio-temporal dynamics. The field measurement is based on the position- and time-dependent energy shift of the probe electron in response to its interaction with the THz field. In this manner, the electron maps the field in both space and time. We further utilize this near field mapping to reconstruct the underlying charge distribution (i.e., the time-evolving carrier density inside the material) that drives the process. The reconstructed charge distribution reveals previously inaccessible aspects of the spatio-temporal evolution of electron and hole densities at the heart of the photo-Dember effect.

We name the demonstrated ultrafast imaging technique *charge dynamics electron microscopy* (CDEM) and develop the theoretical framework describing the electron-field interaction that lies at the core of CDEM, supporting the exploration of ultrafast solid-state charge transport phenomena. This framework encompasses a broad class of interactions between free-electron pulses and charge carriers , where the interaction occurs via the fields produced by the charge dynamics. We show that CDEM can capture instantaneous local electric field distributions with amplitudes as low as $10^3$ V/m in our current setup, and can reach the range of 10 V/m using better electron energy spectrometers that are already commercially available (*39*). CDEM thus opens the door to the observation of a variety of fascinating, previously inaccessible condensed-matter phenomena. For example, we envision the visualization of ultrafast low-temperature superconducting phase transitions, quantum resistivity oscillations, ultrafast metal-insulator transitions, and much more. These effects occur on a sub-picosecond timescale and display nanometer-scale features, making CDEM an ideal approach for their direct observation.

**Results**

Figure 1A describes the interaction mechanism that enables the reconstruction of nanoscale-resolved charge dynamics using free-electron pulses (see ref. (*40*) for further details on the setup). A 60-μm-thick p-type InAs crystal with (111) growth orientation (*40*) is positioned inside the UTEM specimen holder with a cleaved {110} facet facing the pump laser beam. The latter consists of IR pump pulses impinging the InAs crystal at time $t_0$ (Fig. 1A, top panel), inducing transient electron and hole currents. Owing to the photo-Dember effect (*38*), the highly mobile photoexcited electrons rapidly diffuse towards the bulk of the crystal, while the less mobile holes (electron-hole mobility ratio $\mu_e/\mu_h \approx 100$) remain mostly confined near the surface. Diffusion and drift forces lead to picosecond timescale variations in the velocities and spatial densities of the two types of charge carriers. As a result, a THz-frequency electromagnetic pulse is generated and emitted from the crystal.

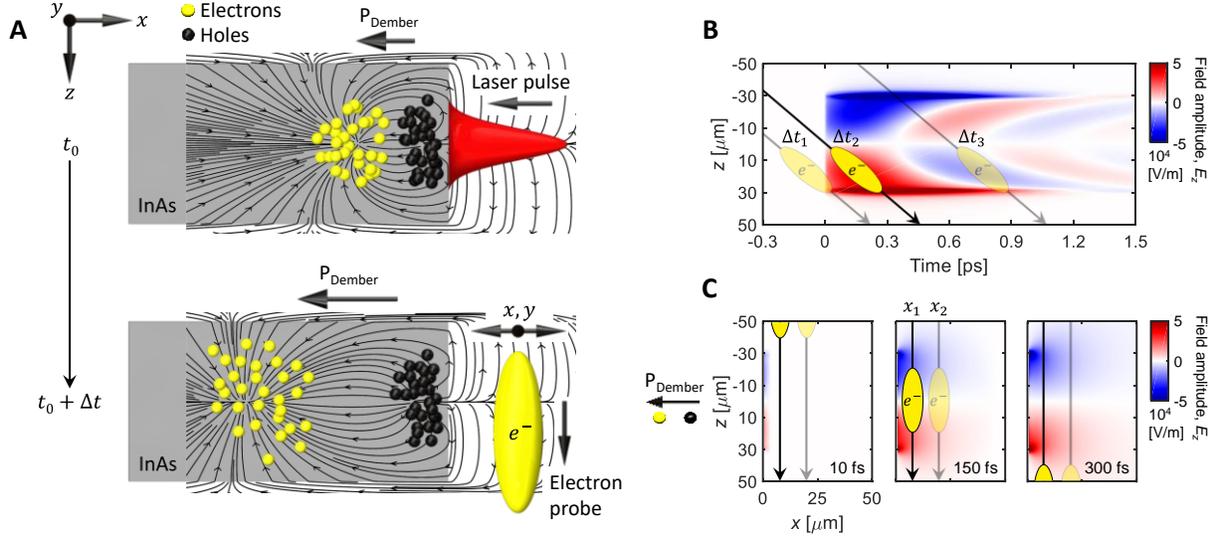

**Fig. 1. Probing near-field THz generation using charge dynamics electron microscopy (CDEM).** (**A**) Illustration of THz pulse generation via the photo-Dember effect and subsequent measurement by a free-electron pulse. An IR laser pulse (red, 40 μm diameter, 50 fs duration, 800 nm central wavelength) impinges on an InAs crystal (gray) from the right at time $t_0$ (top panel). The IR light generates transient electron and hole currents (yellow and black circles, respectively), which lead to the emission of THz-frequency electromagnetic radiation (black electric field lines). A time-delayed electron pulse (yellow ellipse, 200 keV energy) passes by the vicinity of the crystal and undergoes an energy shift (0.1 eV spectrometer resolution) under the influence of the emitted radiation (bottom panel). (**B**) Simulated electric field $z$-component amplitude ($E_z$) versus time and position along the $z$ coordinate (evaluated 1 μm away from the crystal face). (**C**) Simulated $xz$-plane cross sections of $E_z$ at several time delays. Yellow ellipses in (B) and (C) illustrate the electron trajectories (velocity $v = 0.7c$) inside the electric field at different time delays $\Delta t_{1-3}$ (marked in the corresponding panels of Fig. 2B), and at different positions $x_{1-2}$, respectively. Data in (B) and (C) are evaluated at the plane of pump laser incidence and correspond to a pump-laser pulse energy of 10 nJ, with the color bar having 30% saturation for better visibility.

An electron pulse moving along $z$ arrives at a time delay $\Delta t$ after $t_0$ and passes by the vicinity of the pumped crystal face at a lateral distance (impact parameter) $x$ and lateral offset $y$ (Fig. 1A, bottom panel). As illustrated in Figs. 1B and 1C, the THz field changes significantly while the electron passes through the interaction region, resulting in a net energy shift of the electron, measured by using an electron energy-loss spectrometer (EELS). For instance, the electron labeled $\Delta t_2 = 0$ in Fig. 1B, sees no field at $z < 0$ and only positive fields at $z > 0$, and consequently, it decelerates. In contrast, the electron identified by the label $\Delta t_3 = 0.6$ ps, experiences an overall acceleration. Even an electron that arrives at a negative time delay ($\Delta t_1 = -0.24$ ps) can still undergo a nonzero energy shift due to the extended interaction region and electron-pulse duration.

We raster scan the electron position in the $xy$ plane by measuring in scanning-transmission-electron-microscopy (STEM) mode, collecting the electron energy spectrum at each point (see (*40*) for setup details). Each observed average energy shift $\Delta \mathcal{E}$ is determined by an integral of the THz electric field acting on the electron along its trajectory (i.e., the $E_z$ component), thus enabling us to effectively map the phase and amplitude of the associated near field (see theory in (*40*)). We note that the lateral components of the THz near field ($E_x$ and $E_y$) are also accessible through local measurements of the electron deflection in the diffraction plane.

We investigated the electron-field interaction using several sample geometries, and focus the results presented in the figures on a triangle-shaped specimen (Fig. 2), naturally formed when the (111)-grown InAs crystal is cleaved along the {110} facets. This geometry allows us to observe the field pattern emitted from both the excited and unexcited edges, providing valuable information on the spatially-resolved phase of the emitted THz pulse. The internal charge distribution oscillates, forming an *x*-oriented THz dipole having a π-shifted phase on opposite sides of the sample. This behavior is exhibited in the measured spatio-temporal maps (*xy* plane) of the average electron energy shifts $\Delta\mathcal{E}$ around the triangle-shaped InAs crystal, shown in Fig. 2A for various pump-probe time delays. Fig. S3 in (*40*) further shows larger-field-of-view maps taken near a straight-edge crystal (the complete measurement of the field dynamics is presented in Movies S1 and S2). Fig. 2B shows the corresponding simulations, which agree well with the measured data.

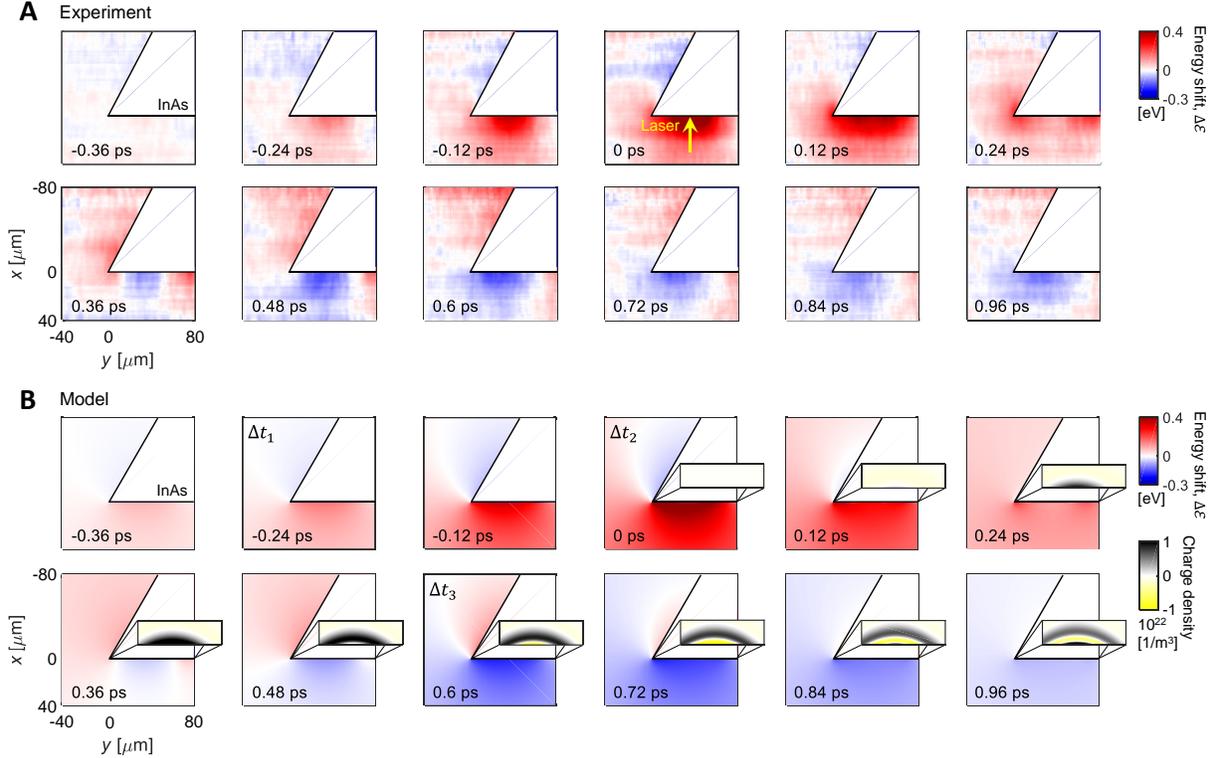

**Fig. 2. Spatio-temporal charge distribution reconstruction via near-field mapping of the THz field.** (**A**) Measured spatio-temporal maps of electron-beam energy shifts. These maps display the mean electron energy shift $\Delta\mathcal{E}$ produced by scanning the electron position in the *xy* plane at various time delays (see labels), thus rendering the THz pulse spatio-temporal dynamics. The pump laser pulse is incident along the *x* axis, normal to the crystal surface, as illustrated by the yellow arrow in the 0 ps panel. The spatial resolution in these measurements of the THz field is dictated by the 1 μm spot size of the probe electron. (**B**) Simulated energy shift $\Delta\mathcal{E}$ corresponding to the experimental data in (A). Insets depict the electron-hole charge distribution reconstructed from the measurements. The cross sections of the charge distribution are taken at the $z = 0$ plane. Insets span 600 nm (80 μm) along *x* (*y*), revealing spatial features of ~100 nm scale in the charge distribution. Labels $\Delta t_{1-3}$ in (B) refer to the time delays indicated in Fig. 1B. All data displayed refer to a pump-laser pulse energy of 10 nJ, with the color bars having 30% saturation for better visibility.

The insets in Fig. 2B show *xy*-plane cross sections of the reconstructed charge distributions at the corresponding points in time. Fig. S4 in (*40*) shows enlarged panels of the charge distributions, as well as *yz*-plane cross sections. Movie S3 combines both cross sections to show the complete reconstructed dynamics. The energy shifts in Fig. 2A are nonzero at negative time delays due to the prolonged interaction with the electron pulse (~350 fs duration), as shown in Fig. 1B, enabling the electron to sense the THz field produced by a pump pulse arriving a fraction of a picosecond earlier. In contrast, the instantaneous charge distributions presented in the insets are zero at negative times, as expected from the short interaction with the excitation pulse (50 fs duration).

From the measured THz field patterns outside the sample, we reconstruct the charge dynamics inside the sample. As with other techniques, such a spatio-temporal reconstruction requires enforcing several constraints on the charge dynamics inside the sample. These constraints enable a unique solution to be obtained from electron measurements outside the crystal volume. In the present instance, sufficient constraints are provided by the use of a hydrodynamic model for the photo-Dember effect (*41*), which prescribes the transient currents inside the sample following photo-excitation by the pump laser. The Poisson equation is subsequently employed to evaluate the associated scalar potential distribution in space and time. This potential is in turn incorporated into a time-dependent Schrödinger equation to find the dynamics of the electron pulse and extract the electron energy spectrum as a function of time delay. The reconstruction of the THz pulse and underlying charge dynamics is done by fitting this theory to the measured five-dimensional data set that is generated by capturing the entire electron energy spectrum as a function of time delay for different *xy* plane positions and pump pulse energies (see theory in (*40*)). Using this approach, we are able to reconstruct the features observed in the experiment with only four fitting parameters, which are found once and for all the recorded data. For this reason, we believe that our simulations adequately reproduce the actual charge distribution generated inside the structure.

The reconstructed net charge distribution penetrates to a depth of ~500 nm inside the crystal bulk (along the *x* axis) and exhibits multiple time-varying annular patterns, visible in both the *xy* and *yz* planes (insets in Fig. 2B, Fig. S4 in (*40*), and Movie S3). Annular patterns in the *yz* plane are produced by local variations in the excited charge carrier density, owing to the Gaussian profile of the pump laser. The varying charge distribution along such a profile causes changes in the back-and-forth (*x* axis) oscillation frequency and intensity at each point along the *yz* plane. These intricate oscillations comprise the microscopic features of the photo-Dember effect in the semiconductor, which result in the emission of a composite THz pulse emerging as the superposition of contributions from different points along the surface. In this manner, CDEM allows us to probe the microscopic details behind the generation of single-cycle THz pulses and to follow their evolution in the near field.

Although InAs-based devices are well-developed for applications, to the best of our knowledge, these spatio-temporal and spectral features have never been observed. Our experiment shows how the emitted-field phase front can be controlled by a combination of sample geometry, local pump laser intensity, and microscopic details at the surface. The intricate dynamics of the charge distribution is manifested by the expanding double-lobe feature exhibited by the emitted THz field, which is clearly discernible in the electron energy-shift maps of Fig. 2 for time delays $\Delta t > 0.24$ ps. Another signature of the charge distribution is the singularity point observed in the energy shift maps around the tip of the triangle-shaped crystal.

The underlying mechanism in CDEM is the inelastic scattering of the free-electron pulse off the charges generated in the probed material, mediated by the THz near field produced by the dynamics of such charges. Thus, each pixel in Fig. 2 depicts the mean electron energy gain or loss extracted from captured time series of electron energy spectra as illustrated in Fig. 3A. Panels in this figure reveal a net energy loss followed by a net gain, with feature strengths depending on the pump-laser pulse energy. Importantly, such features in the electron energy spectrum have not been reported in any previous UTEM experiment. Fig. 3B illustrates the excellent agreement obtained between the measured and simulated mean electron energy shifts extracted from the electron spectra in Fig. 3A. In addition, Fig. 3C compares the measured and simulated peak-to-valley electron energy shifts derived from the data in Fig. 3B for different pump pulse energies, showing a saturation at high pulse energies, which stems from the limited achievable photo-excited charge-carrier density.

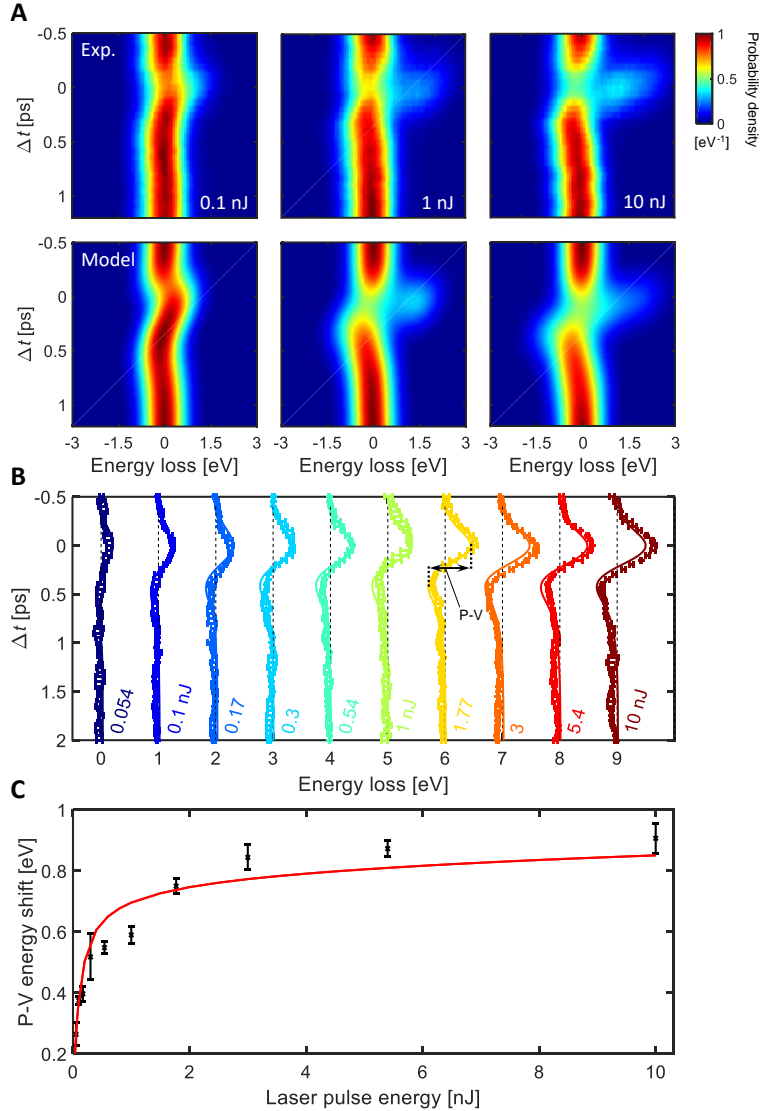

**Fig. 3. Extracting the temporal dynamics of the THz field from the free-electron–field interaction.** (**A**) Measured (top) and simulated (bottom) electron energy spectra versus time delay for several pump-laser pulse energies. Colors indicate the energy probability density, separately normalized at each time delay. Data were collected with the electron beam passing 1 μm away from the crystal face. (**B**) Mean electron energy shift extracted from (A) as a function of time delay. Error bars and lines denote experimental and simulated results, respectively. A horizontal offset is introduced for clarity. (**C**) Peak-to-valley (P-V) electron energy shift (see illustration in (B)) versus pump-laser pulse energy corresponding to the data in (B) (symbols), showing a good agreement with theory (solid curve) (*40*). A common parameter set in the theory matches all data points, for all pulse energies and time delays.

Figure 4 presents an analysis of the spatial dependence of the reconstructed field, showing that the THz spectrum varies with distance from the sample edge, an effect that is unique to near-field measurements. In particular, Fig. 4A depicts the mean electron energy shift as a function of electron-beam position along *x* and time delay for several pump pulse energies, as extracted from the spatio-temporal maps in Fig. 2. The right panels show cross sections along *x* at the time delay of maximal energy shift, together with the corresponding simulated cross sections. Figs. 4B and 4C show the Fourier transform of both the measured and simulated mean electron-energy shift data displayed in Fig. 4A. It is evident that the spectra become narrower as the distance from the crystal grows, a feature that can be traced back to the decay of higher-frequency non-propagating waves in the near field. Interestingly, the experimental spectra in Figs. 4B and 4C show a striking fringe pattern that is not explained by the theory, implying that our experiment is capturing features beyond the hydrodynamic model.

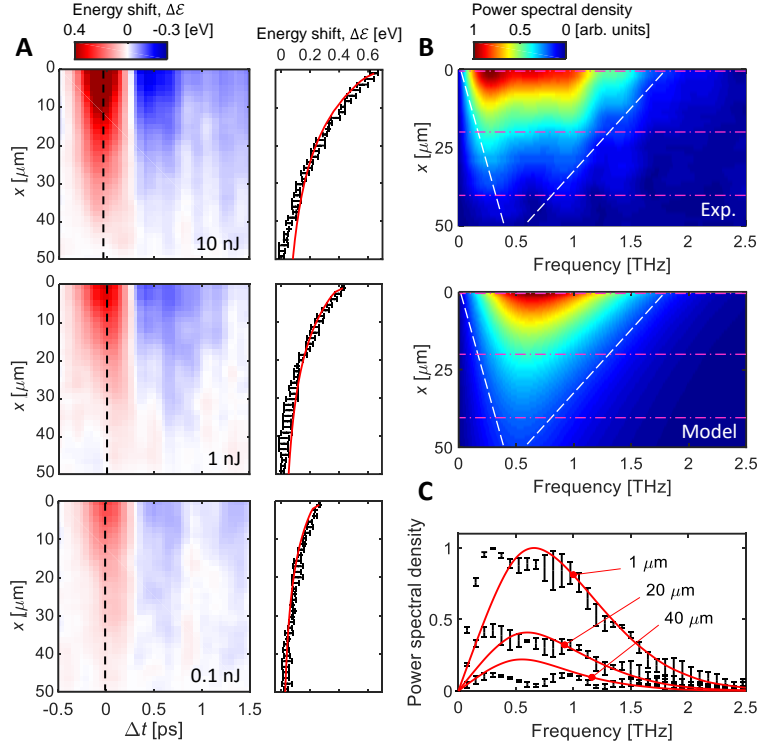

**Fig. 4. Mapping the spectrum of the THz near field.** (**A**) Mean electron energy shift $\Delta\mathcal{E}$ as a function of pump-probe time delay and position along the $x$ coordinate, normal to the crystal face. Data were collected by scanning the electron beam position along the direction of pump-laser incidence $x$, with $x = 0$ at the crystal surface. The color bar has 30% saturation for better visibility. Right panels show a comparison with theory (solid curves) taken at the time delay of maximal energy shift (as indicated by the vertical dashed lines in the left panels). (**B**) Measured (top) and simulated (bottom) power spectral density maps versus position along $x$, as obtained from $\Delta\mathcal{E}$. White dashed lines illustrate a spectra narrowing with increasing distance from the crystal. (**C**) Cross sections of the measured (black error bars) and simulated (red curves) data presented in (B), taken at various $x$ positions (indicated by horizontal magenta lines in (B)). Comparison between theory and experiment shows an agreement in the main profile shape. However, the observed fringe pattern is not explained by the theory, implying that the experiment reveals features beyond the hydrodynamic model. The pump-laser pulse energy in (B) and (C) is 10 nJ.

In perspective, all electron-field interactions harnessed in electron microscopy can be categorized using three parameters, namely, the field cycle, the electron pulse duration, and the interaction duration $L/v$ (determined by the length of the interaction region $L$ and the electron velocity $v$). These parameters allow us to classify all electron-field interactions into three regimes: [1] On the one extreme, for a field cycle much shorter than the electron pulse duration, we are in the realm of photon-induced near-field electron microscopy (PINEM) (*42–49*). [2] The opposite extreme, for a field cycle much longer than both the interaction duration and the electron pulse duration, is the regime of operation of deflectometry, holography, and Lorentz transmission electron microscopy (*26, 50–53*) (more details on the comparison in (*40*)). [3] The intermediate regimes (e.g., for a field cycle longer than the electron pulse duration, yet shorter than the interaction time) are where CDEM belongs.

Our CDEM experiment fundamentally differs from previous PINEM experiments both in its timescale (i.e., we operate in regime [3] rather than [1]) and in its intrinsically nonlinear nature associated with the THz field generation starting from infrared laser light. Namely, in our specific demonstration of CDEM, the electron interacts with a field of a much smaller frequency (THz) than the excitation frequency (optical), created due to the optical nonlinearity induced by the charge dynamics (through the photo-Dember effect). Such nonlinear interactions have not been reported in previous PINEM experiments. CDEM can also occur without nonlinearity, and yet to the best of our knowledge, no realization of CDEM has been reported previously.

Building on previous theoretical developments (*43*, *44*, *47*, *48*, *54*), we formulate a theory for CDEM (*40*). The CDEM theory goes beyond the coherent narrow-band fields commonly employed in PINEM, and can also accommodate a general electromagnetic potential term to account for incoherent and stochastic processes, as often required to characterize complex charge dynamics. This unified quantum theory thus captures both PINEM (regime [1]) and CDEM (regime [3]) on the same footing. The classical limit of this theory is sufficient when the electron pulse duration is substantially shorter than the field cycle, so the electron can be treated as a point particle, as is approximately the case in our experiment.

**Prospects of charge dynamics electron microscopy (CDEM)**

To estimate the prospects of CDEM for the study of other solid-state phenomena, a useful figure of merit is the instantaneous local current $I$ associated with the observed physics (see theory in (*40*)). This current is proportional to the force acting on the probe electron, which in turn produces the electron energy shifts. Based on both measurements and simulations, the lower limit that is observable for this figure of merit using the current setup is $\sim 10^{-2}$ A, corresponding to a $\sim 0.1$ eV energy shift (see theory in (*40*)). This current is equivalent to $\sim 10^5$ electrons undergoing picosecond-scale dynamics. Higher-resolution electron spectrometers that are commercially available can reduce this value by two orders of magnitude (*39*).

These estimates show that CDEM can probe previously inaccessible dynamics, such as the ultrafast nature of superconducting phase transitions (*55*, *56*), recording the formation and resolving the core dynamics of nanoscale vortices in type-II superconductors with nanometer imaging resolution. Here, we provide an estimate for the feasibility of this idea by considering the instantaneous local current $I$ associated with the switching on/off of a single vortex by a femtosecond laser pulse (*57*). As a concrete example, we consider a vortex tube of a single magnetic flux quantum, trapped in a type-II superconductor. Such a vortex is surrounded by a current loop with a typical strength of $10^{-3}$ A (*58*), thus anticipating a 10-meV-scale energy shift (see details in (*40*)), which is measurable via state-of-the-art electron spectrometers (*39*). Moreover, recalling the already known 1-10 ps timescale for the optically-driven phase transition under consideration (*56*), we conclude that the sensitivity of CDEM is sufficient to provide sufficiently-detailed spatio-temporal imaging of this effect.

Looking forward, CDEM experiments utilizing field cycles ranging between ~100 fs and ~10 ps are readily achievable using the current setup. Importantly, this time frame is relevant for many physical phenomena of interest and can be tuned by changing the size of the interaction region or the electron velocity (affecting both the electron pulse duration and its interaction time). Intriguingly, the ~100 fs lower limit could be made three orders of magnitude shorter using attosecond electron pulse trains (*59*).

**Outlook**

In conclusion, CDEM offers a sensitive nonperturbing probe to explore the dynamics of charge carriers inside matter with nanometer-femtosecond spatio-temporal resolution. Our experiment elucidates the intricate electron-hole dynamics inside a photo-excited InAs crystal. The ability to image ultrafast charge dynamics in different materials can assist in future designs of ultrafast nanostructured devices by, for example, optimizing the performance of THz electronics or controlling the beam phase front of metasurface-based THz sources. On a more fundamental level, we expect the CDEM technique to be well-suited to provide quantitative information on the angular distribution of the Fermi velocity of charge carriers in solids – mapping the 3D Fermi surface. This way, CDEM may provide a dynamical nanoscale-resolved alternative to the de Haas – Van Alphen effect (*60–62*).

In general, wherever one has a spatially textured distribution of charge carriers evolving in time and space, like we have in the photo-Dember effect, CDEM can record the associated ultrafast microscopic features. For example, CDEM can facilitate the exploration of hot-carrier dynamics in metals and nano-opto-electronic devices (*63–66*), as well as transport effects related to the magnetoresistance (*67, 68*) and Hall (*69, 70*) effects. These phenomena can be explored in a time-resolved manner to provide new insights into their nature. A similar approach as applied in our experiment could record the dynamics of light-induced superconductivity (*56, 57*), alongside other types of light-induced phase transitions such as those involving charge density waves, the quantum spin Hall effect, and the quantum anomalous Hall effect. Besides phase transitions, one could explore other light-induced phenomena (*71*), including the light-induced anomalous Hall effect, the photovoltaic Hall effect, and ultrafast aspects of Floquet engineering. In all these instances, CDEM grants us access into the microscopic out-of-equilibrium dynamics of charge carriers to map their spatial dependences, as demonstrated in our experiment with the photo-Dember effect.

*We note that a parallel work (Madan et al., "Charge dynamics electron microscopy: nanoscale imaging of femtosecond plasma dynamics") has demonstrated CDEM in a different field, for the visualization of charged plasma dynamics in a UTEM. We submit both papers in parallel aimed at back-to-back publication. The first preliminary results at the basis of both papers were first presented in CLEO 2021 (72, 73).*

**Acknowledgements:** We thank T. Ellenbogen, E. Yalon, and D. Ritter for their support, advice, and helpful discussions. We thank Y. Kauffmann and C. Dickinson for their assistance and advice in STEM operation. We are also grateful to M. Kalina and G. Atiya for assistance in the sample preparation. The experiments were performed on the UTEM of the AdQuanta group of I.K., which is installed in the electron microscopy center (MIKA) of the Department of Materials Science and Engineering at Technion. M.Y. and R.D. are partially supported by the VATAT Quantum Science and Technology scholarship.

**Funding:** The research was supported by the European Research Council (ERC Advanced Grant 789104-eNANO and ERC Starting Grant 851780-NanoEP), the European Union (Horizon 2020 Research and Innovation Program under grant agreement No. 964591 SMART-electron), the Spanish MICINN (PID2020-112625GB-I00 and Severo Ochoa CEX2019-000910-S), the Catalan CERCA Program, and Fundaciós Cellex and Mir-Puig.

**Competing interests:** Authors declare that they have no competing interests.

**Data and materials availability:** All data are available in the main text or the supplementary materials.

**Supplementary Materials**

Materials and Methods

Supplementary Text

Figs. S1 to S4

References (*74*)

Movies S1 to S3